\documentclass[prd,twocolumn,showpacs,preprintnumbers,nofootinbib,amsmath]{revtex4} 
 \usepackage[dvips,final]{graphicx}
  \usepackage{amssymb,amsmath,epsfig,bm,pifont}
   \usepackage{pstricks,pst-node,pst-text,pst-3d}

\newcommand{\MSbar}{\overline{\rm MS}}
\def\1{\hbox{{1}\kern-.25em\hbox{l}}}

\begin{document}
\thispagestyle{empty}
 \date{\today}
\preprint{\hbox{INR-TH-2016-003}}
\preprint{\hbox{USM-TH-341}}

\title{Adler function and
Bjorken polarized sum rule: Perturbation 
expansions in powers of  $SU(N_c)$
conformal anomaly
and studies  of the conformal symmetry limit}.    
\author{Gorazd Cveti\v{c}}
  \email{gorazd.cvetic@usm.cl}
 \affiliation{Physics Department, Universidad T\'ecnica Federico Santa Mar\'{\i}a (UTFSM), Casilla 110-V, Valpara\'iso, Chile\\}
\author{A.~L. Kataev}
  \email{kataev@ms2.inr.ac.ru}
   \affiliation{Institute for Nuclear Research of the Academy of 
Sciences of Russia, 117312, Moscow, Russia;\\
Moscow Institute of Physics and Technology, 141700 Dolgoprudny, 
Moscow Region,Russia}

\begin{abstract}
We consider a new  form of analytical 
perturbation  theory  expansion  in the  
massless $SU(N_c)$ theory, for the nonsinglet part of the  $e^+e^-$-annihilation to 
hadrons Adler function $D^{ns}$  and of  the 
Bjorken sum rule of the polarized lepton-hadron deep-inelastic scattering $C_{ns}^{Bjp}$,
and demonstrate its validity at the 
$O(\alpha_s^4)$-level at least. 
 It is                   a  
 two-fold series in  powers of the conformal anomaly and of
 $SU(N_c)$ coupling  $\alpha_s$.
Explicit expressions are obtained 
for the $\{\beta\}$-expanded perturbation coefficients at 
$O(\alpha_s^4)$  level in  $\overline{\rm MS}$ scheme, for
both considered physical quantities.
Comparisons of the terms in the $\{\beta\}$-expanded
coefficients are made with the corresponding terms
obtained by  using  extra  gluino degrees of freedom, 
or  skeleton-motivated expansion, or  $R_{\delta}$-scheme motivated 
expansion in the Principle of Maximal Conformality.
Relations between terms of the    $\{\beta\}$-expansion
for the $D^{ns}$ and $C_{ns}^{Bjp}$-functions, 
which follow from the conformal symmetry limit and its violation, are 
presented. The relevance to  the possible new  analyses of the experimental 
data for the Adler function and Bjorken sum rule is discussed.
\end{abstract}
\maketitle

It was demonstrated in \cite{Broadhurst:1993ru} that,  
in the $SU(N_c)$ model 
of strong interactions, 
the generalized  $\MSbar$ scheme
Crewther relation between  
the analytically evaluated perturbative
expression 
for the  nonsinglet ($ns$) 
contributions
to the  Adler function 
 and the 
Bjorken sum rule of the polarized lepton-hadron 
deep-inelastic  scattering (DIS)  can be written down as 
\begin{equation}
\label{Eq:1}
D^{ns}(a_s)C_{ns}^{Bjp}(a_s)=1+\Delta_{csb}(a_s)~,
\end{equation} 
 where $\Delta_{csb} \sim a_s^2$,  $a_s \equiv \alpha_s(Q^2)/\pi$,
and $Q^2$ is the physical scale of both $D^{ns}$ and $C_{ns}^{Bjp}$. 
The unity on the rhs  corresponds to the original Crewther 
relation, derived in \cite{Crewther:1972kn} in the 
massless quark-parton model by  applying the  
operator product expansion method  to the   
$\pi^0 \rightarrow \gamma\gamma$ decay   AVV-triangle 
amplitude in  the conformal symmetry (CS) limit.  It   was shown in 
\cite{Broadhurst:1993ru,Gabadadze:1995ei}
that in  $\rm{\overline{MS}}$ scheme the    
CS-breaking (CSB) term  
$\Delta_{csb}$ can be presented  
as a product of the conformal anomaly 
$\beta(a_s)/a_s$ and a polynomial  $P(a_s)$ ($\sim a_s$).
In $\rm{\overline{MS}}$ scheme  the  renormalization group (RG) $\beta$-function is defined as 
\begin{equation}
\nonumber 
\beta(a_s)=\mu^2 \frac{\partial a_s(\mu^2)}{\partial \mu^2}=
-\sum_{j\geq 0}\beta_j a_s(\mu^2)^{j+2} 
\end{equation}
When $\sim a_s^4$
contributions
to $D^{ns}$ and $C_{ns}^{Bjp}$   \cite{Baikov:2010je}
are included, the validity of the
generalized Crewther relation (\ref{Eq:1}) \cite{Broadhurst:1993ru}  at this 
level gets confirmed \cite{Baikov:2010je}.
The $O(a_s^3)$ expression for the $\Delta_{csb}$-term,
fixed in \cite{Broadhurst:1993ru},
is proportional to the two-loop 
expressions of the conformal anomaly, multiplied by 
a polynomial $P(a_s)$ fixed in $\rm{\overline{MS}}$ scheme. 
The term at $a_s^2$ in $P(a_s)$
contains three $SU(N_c)$ group monomials  
$C_F^2$, $C_FC_A$,  $C_FT_F n_f$ of total power 2, composed 
of the Casimir operators $C_F$, $C_A$ and the flavor dependent factor 
$T_F n_f$ (with $T_F=1/2$). 

The expression for  $\Delta_{csb}$ obtained in \cite{Baikov:2010je}
is proportional to the three-loop expression of the conformal anomaly,
multiplied by the same polynomial $P(a_s)$, 
which has the third coefficient 
(at $a_s^3$)
composed of  six $SU(N_c)$ group monomials 
$C_F^3$, $C_F^2C_A$, $C_FC_A^2$, $C_F^2 (T_Fn_f)$, $C_F(T_F n_f)^2$, 
$C_FC_A (T_F n_f)$  of total power 3. 
In \cite{Crewther:1997ux,Braun:2003rp} 
concrete theoretical arguments  were presented showing that
in    $\MSbar$ scheme 
the conformal anomaly  
is factorized in all orders of perturbation theory 
for $\Delta_{csb}$-term in  
Eq.~(\ref{Eq:1}),    
and therefore one should have   
\begin{equation}
\label{gC1}
\Delta_{csb}=\bigg(\frac{\beta(a_s)}{a_s}\bigg)P(a_s) 
=\bigg(\frac{\beta(a_s)}{a_s}\bigg)\sum_{m\geq 1}K_m a_s^m . 
\end{equation}  
In \cite{Kataev:2010du} 
a new form of the  
$\rm{\overline{MS}}$-scheme expression for  
the CSB
term (\ref{gC1}) of the  
generalized  Crewther relation  
was proposed.
It is written down as the two-fold series

\begin{eqnarray}
\nonumber  
&&\Delta_{csb}(a_s)= \sum_{n\geq 1}\bigg(\frac{\beta(a_s)}{a_s}\bigg)^nP_n(a_s) \\ 
\label{gCg}   
&&= \sum_{n\geq 1}\sum_{r\geq1}\bigg(\frac{\beta(a_s)}{a_s}\bigg)^nP_n^{(r)}[k,m]
C_F^kC_A^ma_s^r~~.
\end{eqnarray} 
Here, $r=k+m$ with $k\geq 1$ and $m\geq 0$ , while  
the  coefficients  $P_n^{(r)}[k,m]$ contain rational 
numbers and transcendental Riemann $\zeta_{2l+1}$ functions  with
$l\geq 1$. 
The $SU(N_c)$ 
monomials in Eq.~(\ref{gCg}) {\it do not} contain 
terms proportional to $T_F n_f$, 
in contrast to the less detailed expression in Eq.~(\ref{gC1}) 
where the coefficients $K_m$ ($m \geq 2$) do depend on $T_F n_f$ 
(see \cite{Broadhurst:1993ru,Baikov:2010je} for  explicit
$O(a_s^2)$ and $O(a_s^3)$ results).  
In the 
postulated representation (\ref{gCg}) the dependence 
on $T_Fn_f$ appears in the  powers of $\beta$-function. 
The validity and unambiguity of Eq.~(\ref{gCg}) 
was checked in \cite{Kataev:2010du} 
at the $O(a_s^4)$ level.

One can ask whether   
it is possible to formulate the analogous 
two-fold $\rm{\overline{MS}}$-scheme   
perturbation expansion  for
$D^{ns}$ and $C_{ns}^{Bjp}$ separately, at least at the analytically  
available \cite{Baikov:2010je}  $O(a_s^4)$-level.
Here we present the positive  
answer to this question, and then discuss the main 
consequences of this new QCD resummation 
procedure. In this procedure the  
expansions for $D^{ns}(a_s)$ and $C_{ns}^{Bjp}(a_s)$
take the following form:
\begin{eqnarray}
\label{Dns}
D^{ns}(a_s)&=&1+\sum_{n=0}^{3}\bigg(\frac{\beta(a_s)}{a_s}\bigg)^nD_n(a_s)~,
\\ \label{Cns}
C_{ns}^{Bjp}(a_s)&=&1+\sum_{n=0}^{3}\bigg(\frac{\beta(a_s)}{a_s}\bigg)^nC_n(a_s)~,  
\end{eqnarray} 
where for  $0\leq n \leq 3$,
at the  $O(a_s^4)$-level,
the  polynomials $D_n(a_s)$ and $C_n(a_s)$ are defined as 
\begin{eqnarray}
\label{Dn}
&&D_n(a_s)=\sum_{r= 1}^{4-n} a_s^r \sum_{k=1}^r D_n^{(r)}[k,r-k] C_F^k C_A^{r-k} 
+a_s^4 \delta_{n0} \times
\nonumber 
\\ 
&&
\bigg(D_0^{(4)}[F,A] \frac{d_F^{abcd}d_A^{abcd}}{d_R}+
D_0^{(4)}[F,F] \frac{d_F^{abcd}d_F^{abcd}}{d_R}n_f\bigg),
\\ 
\label{Cn}
&&C_n(a_s)=\sum_{r=1}^{4-n} a_s^r \sum_{k=1}^r C_n^{(r)}[k,r-k] C_F^k C_A^{r-k} 
+a_s^4 \delta_{n0} \times\nonumber 
\\ 
&&
\bigg(C_0^{(4)}[F,A]\frac{d_F^{abcd}d_A^{abcd}}{d_R}+
C_0^{(4)}[F,F]\frac{d_F^{abcd}d_F^{abcd}}{d_R}n_f\bigg).
\end{eqnarray}
The     double sum expressions for 
Eqs.~(\ref{Dn})-(\ref{Cn}) 
are  motivated by the form
for the polynomials $P_n(a_s)$ in Eq.~(\ref{gCg})
introduced in \cite{Kataev:2010du}.
In $SU(N_c)$  
theory and $\rm{\overline{MS}}$ scheme
they  have the unambiguous form  
determined by  the  system of  linear equations, analogous to the 
system presented in \cite{Kataev:2010du}. 
The coefficients with  the  
structures $d_F^{abcd}d_A^{abcd}/d_R$ 
and $d_F^{abcd}d_F^{abcd}/d_R$ 
appear at the 
$O(a_s^4)$ level \cite{Baikov:2010je}. 
These structures  
were defined first in  \cite{vanRitbergen:1997va}, 
where the four-loop coefficient
of the QCD $\beta$-function was  evaluated. 
Since 
$(d_F^{abcd}d_F^{abcd}/d_R)a_s^4$  terms in (\ref{Dn}) and (\ref{Cn})
are proportional to  $T_Fn_f$, which also 
enters the  $\beta_0$-coefficient of the QCD  
$\beta$-function, one 
may propose to move them  
into $D_1(a_s)$ and $C_1(a_s)$-polynomials. 
We will explain  below that    
such a redefinition of    
Eqs.~(\ref{Dns}) and (\ref{Cns}) 
is not supported by  the   QED limit.
Thus  the following   $\rm{\overline{MS}}$-scheme
$O(a_s^{4-n})$ expressions for  
$D_n(a_s)$ ($0 \leq n \leq 3$) are valid: 
\begin{eqnarray}
\nonumber
&&D_0(a_s)=\frac{3}{4}C_Fa_s+\bigg[-\frac{3}{32}C_F^2+\frac{1}{16}C_FC_A\bigg]a_s^2 \\ \nonumber 
&&+\bigg[-\frac{69}{128}C_F^3 -\bigg(\frac{101}{256}-\frac{33}{16}\zeta_3\bigg)C_F^2C_A \\ \nonumber
&&-
\bigg(\frac{53}{192}+\frac{33}{16}\zeta_3\bigg)C_FC_A^2\bigg]a_s^3              
+\bigg[\bigg(\frac{4157}{2048}+\frac{3}{8}\zeta_3\bigg)C_F^4 \\ \label{D0} 
&&
-\bigg(\frac{3509}{1536}
+\frac{73}{128}\zeta_3
+\frac{165}{32}\zeta_5\bigg)C_F^3C_A \\ \nonumber 
&&+\bigg(\frac{9181}{4608}+\frac{299}{128}\zeta_3+\frac{165}{64}\zeta_5\bigg)C_F^2C_A^2 \\ \nonumber 
&&-\bigg(\frac{30863}{36864}+\frac{147}{128}\zeta_3-\frac{165}{64}\zeta_5\bigg)
C_FC_A^3 \\ \nonumber 
&&+   \bigg(\frac{3}{16}-\frac{1}{4}\zeta_3-\frac{5}{4}\zeta_5\bigg)\frac{d_F^{abcd}d_A^{abcd}}{d_R} 
\\ \nonumber 
&&+\bigg(-\frac{13}{16}-\zeta_3+\frac{5}{2}\zeta_5\bigg)\frac{d_F^{abcd}d_F^{abcd}}{d_R}n_f\bigg)\bigg]a_s^4 \ ,
\\ \nonumber
&&D_1(a_s)=\bigg(-\frac{33}{8}+3\zeta_3\bigg)C_Fa_s \\ \nonumber  
&&+\bigg[\bigg(\frac{111}{64}+12\zeta_3-15\zeta_5\bigg)C_F^2 \\ \nonumber 
&&-\bigg(\frac{83}{32}+\frac{5}{4}\zeta_3-\frac{5}{2}\zeta_5\bigg)C_FC_A\bigg]a_s^2 \\ \label{D1}
&&+\bigg[\bigg(\frac{758}{128}+\frac{9}{16}\zeta_3-\frac{165}{2}\zeta_5+\frac{315}{4}\zeta_7\bigg)C_F^3 \\ \nonumber 
&&+\bigg(\frac{3737}{144}-\frac{3433}{64}\zeta_3+\frac{99}{4}\zeta_3^2+\frac{615}{16}\zeta_5-\frac{315}{8}\zeta_7\bigg)C_F^2C_A 
\\ \nonumber 
&&+\bigg(\frac{2695}{384}+\frac{1987}{64}\zeta_3-\frac{99}{4}\zeta_3^2-\frac{175}{32}\zeta_5+\frac{105}{16}\zeta_7\bigg)C_FC_A^2\bigg]a_s^3,
\\ \nonumber 
&&D_2(a_s)= \bigg(\frac{151}{6}-19\zeta_3\bigg) C_Fa_s \\ \nonumber 
&&+\bigg[\bigg(-\frac{4159}{384}-\frac{2997}{16}\zeta_3+27 \zeta_3^2+\frac{375}{2}\zeta_5\bigg)C_F^2 
\\ \label{D2}
&&+\bigg(\frac{14615}{256}+\frac{39}{16}\zeta_3-\frac{9}{2}\zeta_3^2
-\frac{185}{4}\zeta_5\bigg)C_FC_A\bigg]a_s^2 \ , 
\\ \label{D3} 
&&D_3(a_s)= \bigg(-\frac{6131}{36}+\frac{203}{2}\zeta_3+45\zeta_5\bigg)C_Fa_s \ .
\end{eqnarray} 
Analogous expressions  
for the polynomials in  (\ref{Cns})  
read:

\begin{eqnarray}
\nonumber
&&C_0(a_s)=-\frac{3}{4}C_Fa_s+\bigg[\frac{21}{32}C_F^2-\frac{1}{16}C_FC_A\bigg]a_s^2 \\ \nonumber 
&&+\bigg[-\frac{3}{128}C_F^3+\bigg(\frac{125}{256}-\frac{33}{16}\zeta_3\bigg)C_F^2C_A \\ \nonumber 
&&+\bigg(\frac{53}{192}+\frac{33}{16}\zeta_3\bigg)C_FC_A^2\bigg]a_s^3 +\bigg[\bigg(-\frac{4823}{2048}-\frac{3}{8}\zeta_3\bigg)C_F^4 \\ \label{C0}     
&&+\bigg(\frac{605}{384}+\frac{469}{128}\zeta_3+\frac{165}{32}\zeta_5\bigg)C_F^3C_A \\ \nonumber 
&&+\bigg(-\frac{11071}{4608}  -\frac{695}{128}\zeta_3-\frac{165}{64}\zeta_5\bigg)C_F^2C_A^2 \\ \nonumber 
&&+\bigg(\frac{30863}{36864}+\frac{147}{128}\zeta_3-\frac{165}{64}\zeta_5\bigg)C_FC_A^3 \\ \nonumber 
&&+ \bigg(-\frac{3}{16}+\frac{1}{4}\zeta_3+\frac{5}{4}\zeta_5\bigg)\frac{d_F^{abcd}d_A^{abcd}}{d_R} 
\\ \nonumber 
&&+\bigg(\frac{13}{16}+\zeta_3-\frac{5}{2}\zeta_5\bigg)\frac{d_F^{abcd}d_F^{abcd}}{d_R}n_f\bigg)\bigg]a_s^4 \ ,
\\ \nonumber  
&&C_1(a_s)=
\frac{3}{2} C_F a_s +
\bigg[-\bigg(\frac{349}{192}+\frac{5}{4}\zeta_3\bigg)C_F^2 
\\ \nonumber 
&&+\bigg(\frac{155}{96}+\frac{9}{4}\zeta_3-\frac{5}{2}\zeta_5\bigg)C_FC_A\bigg]a_s^2 
\\ \label{C1}
&&+\bigg[\bigg(\frac{997}{384}+\frac{481}{32}\zeta_3-\frac{145}{8}\zeta_5 \bigg)C_F^3 
\\ \nonumber 
&&+\bigg(-\frac{85801}{4608}-\frac{169}{24}\zeta_3+\frac{365}{48}\zeta_5+\frac{105}{4}\zeta_7\bigg)C_F^2C_A 
\\ \nonumber 
&&+\bigg(\frac{931}{768}-\frac{955}{192}\zeta_3-\frac{895}{96}\zeta_5 -\frac{105}{16}\zeta_7\bigg)C_FC_A^2\bigg]a_s^3 \ , 
\\ \nonumber 
&&C_2(a_s)= 
\bigg(-\frac{151}{24}\bigg) C_F a_s 
+\bigg[\bigg(\frac{261}{64}+\frac{87}{8}\zeta_3\bigg)C_F^2 \\ \label{C2} 
&&-\bigg(\frac{3151}{256}+\frac{43}{16}\zeta_3+\frac{3}{2}\zeta_3^2-\frac{15}{4}\zeta_5\bigg)C_FC_A\bigg]a_s^2 \ ,
\\ \label{C3} 
&&C_3(a_s)= 
\frac{605}{36} C_F a_s \ .
\end{eqnarray} 

The  singlet  ($si$)
corrections to the Adler function and to  the Bjorken sum rule      
should be considered separately
(see \cite{Brodsky:2013vpa,Kataev:2014jba}). 
In the Adler function they 
appear first  in  $O(a_s^3)$  
\cite{Gorishnii:1990vf,Surguladze:1990tg,Chetyrkin:1996ez} and are 
known up to $a_s^4$ \cite{Baikov:2012zn}.
For the Bjorken sum rule they start to  contribute at 
$O(a_s^4)$  \cite{Larin:2013yba,Baikov:2015tea}.  For   $n_f=3, 6$
the $si$  contributions to both quantities are  equal 
to zero. For the cases of $n_f=4, 5$ they  
are significantly   smaller then the
$ns$-effects. 

We explain how the results (\ref{D0})-(\ref{D3}) and (\ref{C0})-(\ref{C3}) 
were obtained.
The coefficients $\beta_0$, $\beta_1$, $\beta_2$ 
of the RG $\beta$-function  
on the r.h.s.
 of (\ref{Dns}) and  (\ref{Cns})
are known in terms of powers of  $C_F$, $C_A$ and $T_F n_f$.  
The  $\beta_0$-term  was evaluated in 
\cite{Gross:1973id,Politzer:1973fx}, $\beta_1$
in \cite{Caswell:1974gg,Jones:1974mm,Egorian:1978zx}, 
$\beta_2$ in $\MSbar$ 
in 
\cite{Tarasov:1980au,Larin:1993tp}. 
To determine the
coefficients $D_n^{(r)}[k,m]$ , $C_n^{(r)}[k,m]$ in 
(\ref{Dn}) and (\ref{Cn})  
the l.h.s.  of 
Eqs.~(\ref{Dns})-(\ref{Cns}) 
is expressed as 

\begin{eqnarray}
\label{Das}
D^{ns}(a_s)=1+d_1a_s+d_2a_s^2+d_3a_s^3+d_4a_s^4+O(a_s^5), \\ 
\label{Cbjas}
C_{ns}^{Bjp}(a_s)=1+c_1a_s+c_2a_s^2+c_3a_s^3+c_4a_s^4+O(a_s^5),
\end{eqnarray} 
and the $\rm{\overline{MS}}$-coefficients expanded in color structures   of  
the $SU(N_c)$ group.    
The  coefficients   
$d_1$-$d_4$ are known 
from 
the works 
\cite{Appelquist:1973uz, Chetyrkin:1979bj}~,~\cite{Gorishnii:1990vf}
and \cite{Baikov:2010je}, correspondingly, 
while $c_1$-$c_4$  
were evaluated in 
\cite{Kodaira:1978sh,Gorishnii:1985xm}~,~\cite{Larin:1991tj} and 
\cite{Baikov:2010je}, respectively.
 Following the logic of \cite{Kataev:2010du}, we used in
Eq.~(\ref{Dns}) on the l.h.s.  the expansion (\ref{Das}), and 
on the r.h.s.  the expansions (\ref{Dn}) for $D_n(a_s)$ 
and the expansions 
in terms of $C_F$, $C_A$
and $T_F n_f$ of the RG $\beta$-function coefficients. 
Equating
the expressions at
all monomials in $C_F$, $C_A$ and $T_F n_f$  
at each power of $a_s$ on both sides of Eq.~(\ref{Dns})
leads to a complete system of 22  linear equations, analogous to the 
(smaller) system in \cite{Kataev:2010du}. 
Its unique solution 
determines the  polynomials $D_n(a_s)$ ($0 \leq n \leq 3$)
in Eqs.~(\ref{D0})-(\ref{D3}).
To get the results (\ref{C0})-(\ref{C3}), 
the analogous procedure  
is applied 
to $C_{ns}^{Bjp}(a_s)$.
As a cross-check  we reproduced the results of \cite{Kataev:2010du} 
for Eq. (\ref{gCg}).\footnote{
Note that Eq.~(15) in \cite{Kataev:2010du} contains 
a misprint. The $C_FC_A^2a_s^3$ contribution to $P_1(a_s)$, defined in 
Eq.~(\ref{gCg}), should contain an extra $3/4$ factor.} 

In the CS limit, i.e.,  when $\beta \mapsto 0$ in 
Eqs.~(\ref{Dns}) and (\ref{Cns}), we get  
(cf.~an analogous identity in \cite{Kataev:2014jba}):
\begin{equation}
\label{CrQCD}
(1+D_0(a_s(Q^2))\times(1+C_0(a_s(Q^2))=1,
\end {equation}
where $D_0(a_s)$ and $C_0(a_s)$ are given in (\ref{D0}) and (\ref{C0}).  
The terms proportional to
$d_F^{abcd}d_A^{abcd}/d_R$ and $n_f d_F^{abcd}d_F^{abcd}/d_R$,
in (\ref{D0}) and (\ref{C0}), cancel out in Eq.~(\ref{CrQCD}).
This identity is an extension of the Crewther relation, 
derived in \cite{Crewther:1972kn} in the Born approximation.
   
We can now fix the 
$\{\beta\}$-expansion structure  (proposed in \cite{Mikhailov:2004iq})  of
the coefficients $d_j$ of $D^{ns}$
\begin{eqnarray}
\label{d2}
&&d_1=d_1[0]~,~d_2= \beta_0 d_2[1]+ d_2[0]~, \\  \label{d3} 
&&d_3=\beta_0^2d_3[2] + \beta_1 d_3[0,1]+ \beta_0 d_3[1] + d_3[0]~, \\ 
\label{d4}
&&d_4=\beta_0^3 d_4[3] + \beta_1 \beta_0 d_4[1,1] + \beta_2 d_4[0,0,1] \\ 
\nonumber  
&&+ \beta_0^2 d_4[2] + \beta_1  d_4[0,1] + \beta_0\,d_4[1] +d_4[0] \ .
\end{eqnarray}                                                 
In \cite{Mikhailov:2004iq}, this was performed
up to  $O(a_s^3)$-level only,  with
the $SU(N_c)$ model supplemented by a multiplet of 
gluino degrees of freedom of SUSY QCD.\footnote{
The validity of the 
$O(a_s^3)$
of the  $\{\beta\}$-expansion  
results  of \cite{Mikhailov:2004iq} was 
confirmed recently in \cite{Ma:2015dxa}.}. 

Applying the  two-fold expansion (\ref{Dns})
and the $SU(N_c)$ results
(\ref{D0})-(\ref{D3}), 
we  obtain all 
$\{\beta\}$-expanded terms in  $d_2$, $d_3$ and even  $d_4$ 
$\MSbar$-scheme coefficients: 
\begin{eqnarray}
\nonumber 
&&d_1[0]=\frac{3}{4}C_F~,~d_2[0]=\bigg(-\frac{3}{32}C_F^2+\frac{1}{16}C_FC_A
\bigg)~, \\ \nonumber  
&&d_2[1]=\bigg(\frac{33}{8}-3\zeta_3\bigg)C_F~,~  
d_3[0]=\bigg[-\frac{69}{128}C_F^3  
\\ \nonumber 
&&-\bigg(\frac{101}{256}-
\frac{33}{16}\zeta_3\bigg)C_F^2C_A 
-\bigg(\frac{53}{192}+\frac{33}{16}\zeta_3\bigg)C_FC_A^2\bigg] \ , 
\\ \nonumber  
&&d_3[1]=\bigg[\bigg(-\frac{111}{64}-12\zeta_3+15\zeta_5\bigg)C_F^2  \\ \nonumber &&+\bigg(\frac{83}{32}+\frac{5}{4}\zeta_3-\frac{5}{2}\zeta_5\bigg)C_FC_A\bigg] \ ,
\\ \nonumber
&&d_3[0,1]=\bigg(\frac{33}{8}-3\zeta_3\bigg)C_F,~ 
d_3[2]=\bigg(\frac{151}{6}-19\zeta_3\bigg)C_F, \\ \nonumber
&&d_4[0]=\bigg[\bigg(\frac{4157}{2048}+\frac{3}{8}\zeta_3\bigg)C_F^4 
\\ \label{d4new} 
&&-\bigg(\frac{3509}{1536}+\frac{73}{128} \zeta_3+\frac{165}{32}\zeta_5\bigg)C_F^3C_A 
\\ 
\nonumber  
&&+\bigg(\frac{9181}{4608}+\frac{299}{128}\zeta_3+\frac{165}{64}\zeta_5\bigg)C_F^2C_A^2 \\ \nonumber 
&&-\bigg(\frac{30863}{36864}+\frac{147}{128}\zeta_3
-\frac{165}{64}\zeta_5\bigg)C_FC_A^3 \\ \nonumber   
&&+ \bigg(\frac{3}{16}-\frac{1}{4}\zeta_3-\frac{5}{4}\zeta_5\bigg)\frac{d_F^{abcd}d_A^{abcd}}{d_R}  
\\ \nonumber   
&& +\bigg(-\frac{13}{16}-\zeta_3+\frac{5}{2}\zeta_5\bigg)\frac{d_F^{abcd}d_F^{abcd}}{d_R}n_f  \ ,
\\ \nonumber 
&&d_4[1]=\bigg(-\frac{785}{128}-\frac{9}{16}\zeta_3+\frac{165}{2}\zeta_5-
\frac{315}{4}\zeta_7\bigg)C_F^3 \\ \nonumber 
&&-\bigg(\frac{3737}{144}-\frac{3433}{64}\zeta_3
+\frac{99}{4}\zeta_3^2 +\frac{615}{16}\zeta_5-\frac{315}{8}\zeta_7\bigg)C_F^2C_A \\ \nonumber 
&&-\bigg(\frac{2695}{384}+\frac{1987}{64}\zeta_3-\frac{99}{4}\zeta_3^2
-\frac{175}{32}\zeta_5+\frac{105}{16}\zeta_7\bigg)C_FC_A^2 \ ,
\\ \nonumber 
&&d_4[0,1]=\bigg[\bigg(-\frac{111}{64}-12\zeta_3+15\zeta_5\bigg)C_F^2 
\\ \nonumber
&&+\bigg(\frac{83}{32}+\frac{5}{4}\zeta_3-\frac{5}{2}\zeta_5\bigg)C_FC_A\bigg] \ ,
\\ 
\nonumber 
&&d_4[2]=\bigg(-\frac{4159}{384}-\frac{2997}{16}\zeta_3+27\zeta_3^2
+\frac{375}{2}\zeta_5\bigg)C_F^2 
\\ \nonumber  
&&+\bigg(\frac{14615}{256}+\frac{39}{16}\zeta_3-\frac{9}{2}\zeta_3^2 -\frac{185}{4}\zeta_5\bigg)C_FC_A \ ,
\\ \nonumber 
&&d_4[0,0,1]=\bigg(\frac{33}{8}-3\zeta_3\bigg)C_F~,
\\ \nonumber 
&&d_4[1,1]=\bigg(\frac{151}{3}-38\zeta_3\bigg)C_F \ , 
\\ \nonumber  
&&d_4[3]=
\bigg(\frac{6131}{36}-\frac{203}{2}\zeta_3-45\zeta_5\bigg)C_F \ .
\end{eqnarray}
The  $\{\beta\}$-expanded  
coefficients  of $C^{Bjp}_{ns}$  
have  the same structure as Eqs.~(\ref{d2})-(\ref{d4}):   
\begin{eqnarray}
\label{c2}
&&c_1=c_1[0]~~,~~ c_2= \beta_0 c_2[1]+ c_2[0]~, \\ \label{c3} 
&&c_3=\beta_0^2c_3[2] + \beta_1 c_3[0,1]+ \beta_0 c_3[1] + c_3[0]~, \\ 
\label{c4}
&&c_4=\beta_0^3 c_4[3] + \beta_1 \beta_0 c_4[1,1] + \beta_2 c_4[0,0,1] \\ 
\nonumber  
&&+ \beta_0^2 c_4[2] + \beta_1  c_4[0,1] + \beta_0\,c_4[1] +c_4[0].
\end{eqnarray}
Using the two-fold series (\ref{Cns}) and Eqs.~(\ref{C0})-(\ref{C3}),  we get
\begin{eqnarray}
\nonumber 
&&c_1[0]=-\frac{3}{4}C_F~,~~c_2[0]=\bigg(\frac{21}{32}C_F^2-\frac{1}{16}C_FC_A
\bigg) \ ,
\\ \nonumber 
&&c_2[1]=-\frac{3}{2}C_F~,~~c_3[0]=\bigg[-\frac{3}{128}C_F^3 \\ \nonumber 
&&+\bigg(\frac{125}{256}-\frac{33}{16}\zeta_3\bigg)C_F^2C_A+\bigg(\frac{53}{192}
+\frac{33}{16}\zeta_3\bigg)C_FC_A^2\bigg]  \ ,
\\ \nonumber 
&&c_3[1]=\bigg(\frac{349}{192}+\frac{5}{4}\zeta_3\bigg)C_F^2 
-\bigg(\frac{155}{96}+\frac{9}{4}\zeta_3-\frac{5}{2}\zeta_5\bigg)C_FC_A, 
\\ \nonumber 
&&c_3[0,1]= -\frac{3}{2}C_F ~,~ c_3[2]=-\frac{115}{24}C_F \ , \\ \nonumber 
&&c_4[0]=\bigg[\bigg(-\frac{4823}{2048}-\frac{3}{8}\zeta_3\bigg)C_F^4 
\\ \label{c4new}
&&+\bigg(\frac{605}{384}+\frac{469}{128}\zeta_3+\frac{165}{32}\zeta_5\bigg)C_F^3C_A \\ \nonumber  
&&+\bigg(-\frac{11071}{4608}-\frac{695}{128}\zeta_3-\frac{165}{64}\zeta_5\bigg)C_F^2C_A^2 \\ \nonumber 
&&+\bigg(\frac{30863}{36864}+\frac{147}{128}\zeta_3
-\frac{165}{64}\zeta_5\bigg)C_FC_A^3 \\ \nonumber 
&&+ \bigg(-\frac{3}{16}+\frac{1}{4}\zeta_3+\frac{5}{4}\zeta_5\bigg)\frac{d_F^{abcd}d_A^{abcd}}{d_R}  
\\ \nonumber 
&& +\bigg(\frac{13}{16}+\zeta_3-\frac{5}{2}\zeta_5\bigg)\frac{d_F^{abcd}d_F^{abcd}}{d_R}n_f\bigg] \ ,
\\ \nonumber  
&&c_4[1]=\bigg[\bigg(-\frac{997}{384}-\frac{481}{32}\zeta_3+\frac{145}{8}\zeta_5\bigg)C_F^3 \\ \nonumber 
&&+\bigg(\frac{85801}{4608}+\frac{169}{24}\zeta_3
-\frac{365}{48}\zeta_5-\frac{105}{4}\zeta_7\bigg)C_F^2C_A \\ \nonumber 
&&-\bigg(\frac{931}{768}-\frac{955}{192}\zeta_3
-\frac{895}{96}\zeta_5-\frac{105}{16}\zeta_7\bigg)C_FC_A^2\bigg] \ , 
\\ \nonumber  
&&c_4[0,1]=\bigg(\frac{349}{192}+\frac{5}{4}\zeta_3\bigg)C_F^2  
\\ \nonumber 
&&-\bigg(\frac{155}{96}+\frac{9}{4}\zeta_3-\frac{5}{2}\zeta_5\bigg)C_FC_A,  
\\ \nonumber  
&&c_4[2]=\bigg[\bigg(\frac{261}{64}+\frac{87}{8}\zeta_3\bigg)C_F^2 \\ \nonumber 
&&-\bigg(\frac{3151}{256}+\frac{43}{16}\zeta_3+\frac{3}{2}\zeta_3^2
-\frac{15}{4}\zeta_5\bigg)C_FC_A \bigg]
 \\ \nonumber 
&&c_4[0,0,1]= -\frac{3}{2}C_F~,~c_4[1,1]=-\frac{115}{12}C_F ~,
\\ \nonumber 
&&c_4[3]=-\frac{605}{36}C_F.  
\end{eqnarray}
Note that specific contributions to  $d_3$ and $c_3$  
differ  from those given in 
\cite{Mikhailov:2004iq,Kataev:2010du,Kataev:2014jba}.
The results for the  $\{\beta\}$-expansion of $d_4$ and $c_4$ are new. 

As mentioned,  formally  it is  possible to rewrite  the 
$a_s^4\delta_{n0}(d_F^{abcd}d_F^{abcd}/d_R)n_f$ contribution to   
Eqs.~(\ref{Dn}) and (\ref{Cn})  
\begin{eqnarray}
\nonumber
&&a_s^4\delta_{n0}D_0^{(4)}[F,F]\frac{d_F^{abcd}d_F^{abcd}}{d_R}n_f \mapsto
\\ 
&&a_s^4
\big(\delta_{n0}\frac{11C_A}{4T_F}D_0^{(4)}[F,F] 
+\delta_{n1}\frac{3}{T_F}D_1^{(4)}[F,F]\big)\frac{d_F^{abcd}d_F^{abcd}}{d_R},
\nonumber\\
\label{version} 
\end{eqnarray}
where $D_0^{(4)}[F,F]=D_1^{(4)}[F,F]$.
This leads to   rearrangements of the $a_s^4(d_F^{abcd}d_F^{abcd}/d_R)$ 
terms in  
(\ref{D0}) between the 
$a_s^4$ terms of Eqs.~(\ref{D0}) and (\ref{D1}), and to  the
redefinitions   of the terms   $d_4[0]$ and $d_4[1]$ in the 
$\{\beta\}$-expansion of the coefficient $d_4$  
\begin{eqnarray}
\nonumber 
d_4^{mod}[0]&=&d_4[0] -D_0^{(4)}[F,F]\frac{d_F^{abcd}d_F^{abcd}}{d_R}n_f \\ 
\label{4mod0}
&+& \frac{11}{4}D_0^{(4)}[F,F]\frac{C_Ad_F^{abcd}d_F^{abcd}}{T_Fd_R} \ , \\ 
\label{4mod1}
d_4^{mod}[1]&=&d_4[1] - 3D_1^{(4)}[F,F]\frac{d_F^{abcd}d_F^{abcd}}{T_Fd_R} \ ,      
\end{eqnarray}
where  $D_0^{(4)}[F,F]$=$D_1^{(4)}[F,F]$=$(-13/16-\zeta_3+5\zeta_5/2)$.
This gives $n_f$-independent term $d_4^{mod}[0]$.  However, 
this rearrangement   is not supported 
by the QED limit, which should be valid in the case 
of  theoretically self-consistent definition  
of the new resummed representations of    Eq.~(\ref{Dns}) 
and of the  related 
$\{\beta\}$-expanded expressions for the coefficients $d_i$.
This  QED limit is  realized by fixing $C_A$=0, $T_F=1$, 
$d_F^{abcd}d_F^{abcd}/d_R=1$ and $n_f=N$, 
where $N$ is the number of leptons. 
In QED the  remaining   $D_0^{(4)}[F,F]$-contribution      
arises from the five-loop Feynman diagram with 
light-by-light scattering internal subgraph, contributing to the photon
vacuum polarization function. However, this subgraph is 
convergent and does not give  extra $\beta_0$-dependent 
(or $N$-dependent) contribution 
to the coefficient $d_4$.
Therefore we prefer the definitions of Eqs.~(\ref{Dn}) and 
(\ref{Cn}) without applying to them 
the rearrangements  of Eq.~(\ref{version}).    
Note  also  that  $d_F^{abcd}d_F^{abcd}$ structure is contributing 
the $n_f^2$ part of the four-loop coefficient of the RG $\beta$-function 
in $SU(N_c)$ theory \cite{vanRitbergen:1997va,Czakon:2004bu}, 
which is manifesting itself in
Eqs.~(\ref{Dn}) and (\ref{Cn}) only  starting from the
unknown $a_s^5$ corrections. This
is an extra argument which disfavours
the $a_s^4$ rearrangements
(\ref{4mod0})-(\ref{4mod1}).  
 
We  now discuss common features and differences between 
the results for the 
$\{\beta\}$-expanded  coefficients $d_i$ and $c_i$,  
obtained with
various perturbative approaches for  
$D^{ns}(a_s)$ and $C_{ns}^{Bjp}(a_s)$.
Consider the $\{\beta\}$-expansion results obtained 
with:
$\rm(I)$
the $\{\beta\}$-expansion formalism  
\cite{Mikhailov:2004iq} (cf.~also \cite{Kataev:2010du,Kataev:2014jba});~ 
$\rm(II)$  the  $\{\beta\}$-expansion formalism
based on the resummed  
Eqs.~(\ref{Dns}) and (\ref{Cns})
proposed here;~
$\rm(III)$ skeleton-motivated expansion 
\cite{Cvetic:2006gc} (Section IV there);~ 
$\rm(IV)$ $R_{\delta}$-scheme motivated expansion
of the  Principle of Maximal Conformality
\cite{Brodsky:2013vpa}, 
\cite{Mojaza:2012mf,Wu:2013ei}.\\
In all 
four approaches
the leading $\beta_0$-terms
$d_n[n-1] \beta_0^{n-1}$ (and $c_n[n-1]\beta_0^{n-1}$),
coincide. They coincide also with   
the  leading  $\beta_0$-terms in the 
$\beta_0$-expansion of \cite{LovettTurner:1995ti}, and with the 
corresponding terms of  the 
large $\beta_0$-extension  \cite{Beneke:1994qe}
of Brodsky-Lepage-Mackenzie (BLM) approach \cite{Brodsky:1982gc}.
This feature is a consequence of a direct relation 
of these terms
with the  renormalon contributions \cite{Broadhurst:1993ru}
to the expressions 
for $D^{ns}(a_s)$ and $C^{Bjp}_{ns}(a_s)$.

Further, the approaches $\rm{I}$-$\rm{IV}$ 
generate the same structure of $\{\beta\}$-expansion
of the coefficients $d_i$ and $c_i$,
cf.~Eqs.~(\ref{d2})-(\ref{d4}) and Eqs.~(\ref{c2})-(\ref{c4}).

However, specific coefficients  in the $\{\beta\}$-expanded 
expressions of $d_3$ and $c_3$ obtained here
do not coincide with those obtained in 
\cite{Mikhailov:2004iq,Kataev:2010du,Kataev:2014jba}.
Only the $C_F^3$-terms coincide. The latter is 
a consequence of realization of the CS  and therefore of the 
Crewther relation 
of Eq.~(\ref{CrQCD}) in the perturbative  quenched QED approximation 
(cf.~discussions in \cite{Kataev:2013vua,Adler:1973kz}).  
The   
analytical expressions for the  $C_F^2C_A$, $C_FC_A^2$ 
contributions  to 
$\beta_i$-independent $d_3[0]$ and $c_3[0]$
components of 
$d_3$ and $c_3$ in (\ref{d3}) and (\ref{c3}), 
and for the terms  $d_3[0,1]$, $d_3[1]$, and $c_3[0,1]$, $c_3[1]$,
differ from  the  expressions obtained in
\cite{Mikhailov:2004iq} [cf.~Eqs.~(\ref{d4new}) and (\ref{c4new}) with the
corresponding results in \cite{Mikhailov:2004iq,Kataev:2010du,Kataev:2014jba}].
This difference arises because the $\{\beta\}$-expansion 
formalism in \cite{Mikhailov:2004iq,Kataev:2014jba} was performed
in a gauge model which, in addition to $SU(N_c)$, 
contains a gluino multiplet, while 
the QCD  results obtained here,
including the identities 
\begin{eqnarray}
\label{id}
d_2[1]&=&d_3[0,1]=d_4[0,0,1] =\left(\frac{33}{8}-3\zeta_3\right) ,
\\ 
c_2[1]&=&c_3[0,1]=c_4[0,0,1] = \left(-\frac{3}{2}C_F \right) ,
\label{ic}
\end{eqnarray}
use  
special resummation approach of Eqs.~(\ref{Dns}) and (\ref{Cns}). 
This approach is unambiguously defined up to  
$O(a_s^4)$ within the $SU(N_c)$ gauge model,
while the approach of \cite{Mikhailov:2004iq,Kataev:2010du,Kataev:2014jba}
is at the moment defined only up to $O(a_s^3)$.
We note
that the identities (\ref{id})-(\ref{ic})
hold in the resummation approaches $\rm(III)$ 
and $\rm(IV)$ as well.
Moreover, it turns out that at $O(a_s^4)$ level 
the $\{\beta\}$-expansions of perturbation coefficients in the
approaches  $\rm(III)$ and $\rm(IV)$, 
i.e., in the skeleton method \cite{Cvetic:2006gc}, 
and $R_{\delta}$-scheme method \cite{Mojaza:2012mf},\cite{Brodsky:2013vpa}
are similar to each other
\footnote{The details of these formulations and comparisons
will be considered elsewhere.}.   
The relations between these methods III and IV and the method
developed here reside in a 
careful application of the RG method (for the stages of its development see  
\cite{BSHbook}).

Note that in this work the concept of CS  and the effects of
CSB were essential to obtain new analytical  results
of  Eqs.~(\ref{d4new}) and (\ref{c4new}).   
These  concepts allowed   to  derive  in  \cite{Kataev:2010du,Kataev:2014jba}
the number of  relations from formulated in  \cite{Kataev:2010du}  
Eq.~(\ref{gCg}). Therefore, the results obtained above  satisfy 
them:
\begin{eqnarray}
\nonumber 
&&c_3[0]+d_3[0]=2d_1d_2[0]-d_1^3 
=-\frac{9}{16}  C_F^3 + \frac{3}{32} C_F^2 C_A, \\ \nonumber 
&&c_4[0]+d_4[0]=2d_1d_3[0]-3d_1^2d_2[0]+d_2[0]^2+d_1^4  = \\ \label{numb2}
&& -\frac{333}{1024} C_F^4 + 
\left( - \frac{363}{512} + \frac{99}{32} \zeta_3 \right) C_F^3 C_A
\\ \nonumber
&&- \left( \frac{105}{256} +  \frac{99}{32} \zeta_3 \right) C_F^2 C_A^2, 
\\ \nonumber 
&&c_2[1]+d_2[1]=c_3[0,1]+d_3[0,1]=c_4[0,0,1]+d_4[0,0,1] \\  \nonumber
&&=\bigg(\frac{21}{8}-3\zeta_3\bigg)C_F, \\ \nonumber
&&c_3[1]+d_3[1]+d_1(c_2[1]-d_2[1]) \\ \label{numb3}
&&=c_4[0,1]+d_4[0,1]+d_1(c_3[0,1]-d_3[0,1])\\ \nonumber
&&=-\bigg(\frac{397}{96}+\frac{17}{2}\zeta_3-15\zeta_5\bigg)C_F^2 
+\bigg(\frac{47}{48}-\zeta_3\bigg)C_FC_A~.
\end{eqnarray}
We note,  that these relations and expressions are model-independent
and scheme-independent. They
are also valid in the approaches III and IV. 
These expressions may be used as a check if  the 
$\{\beta\}$-expansion formalism in QCD with additional 
degrees of freedom  \cite{Mikhailov:2004iq}, also  considered in
~\cite{Kataev:2010du,Kataev:2014jba},  
is extended to $d_4$ and $c_4$.

The results obtained in this work may be used in
future  phenomenologically oriented
studies of various resummation procedures and of their 
relations to generalizations of the BLM approach,  
related to  Principle of Maximal Conformality \cite{Brodsky:2013vpa,Wu:2013ei},
i.e. the ones considered recently in  \cite{Kataev:2014jba,Ma:2015dxa},  
and to the skeleton-motivated approach \cite{Cvetic:2006gc}.
 Here we comment on 
a link of our studies with a specific result  of  the 
generalized BLM method, written 
down in the form of commensurate scale relations
\cite{Brodsky:1994eh}, namely with the expression \cite{Brodsky:1995tb}
\begin{equation}
\label{BKKL}
\big(1+a_s^{Dns}(Q^{*}_{Dns})\big)\big(1+a_s^{Bns}(Q^{*}_{Bns})\big)=1~~~. 
\end{equation} 
This expression follows from the  generalized Crewther 
relation of  \cite{Broadhurst:1993ru} after defining the effective charges 
of the non-singlet contributions to the 
Adler function  and to  the Bjorken polarized sum rule 
using the 
effective-charge approach \cite{Grunberg:1982fw} and  absorbing  
the  $\beta$-function dependent terms into the effective  
scales  of the running  effective charges
$a_s^{Dns}$ and $a_s^{Bns}$. The expression (\ref{BKKL}) is similar 
in its form to the QCD relation (\ref{CrQCD})
derived here  in the conformal invariant limit.
The CSB effects are manifested in Eq.~(\ref{BKKL}) in the (different)
values of the effective scales $Q^{*}_{Dns}$ and  $Q^{*}_{Bns}$. 
The empirical, experimentally-motivated, consideration for the importance of  
these CSB effects at sufficiently  high energies was presented in  
\cite{Brodsky:1995tb}. 

We hope that the representation for the Adler function obtained here
  can be used in a more detailed comparison with the expression for the
  Adler function obtained in \cite{Eidelman:1998vc} 
from the available data 
for the  $e^+e^-$-annihilation 
to hadrons total cross-section.
Analogous comparison can be performed for the obtained Bjorken sum rule
representation with the Bjorken sum rule most recent data,  determined in
\cite{Deur:2014vea} for the $Q^2\leq 4.8$ $\rm{GeV^2}$ region.

\vspace{0.5cm} 

\section*{Acknowledgments}
One of us (A.K.) is  grateful  to   D.J.~Broadhurst and C.J.~Maxwell 
for discussions of the ideas which 
motivated this work.
Both of us would like to thank S.V.~Mikhailov for his interest in 
the results of our studies and useful discussions.  
Useful comments of R.J.~Crewther and S.J.~Brodsky to the 
first version of this work are gratefully acknowledged.
We dedicate this  manusript to D.V. Shirkov. 
The work of A.K.
was supported by grant No.~14-22-00161 of the
Russian Science Foundation. 
The work of G.C. was supported by FONDECYT (Chile) Grant No.~1130599
and by UTFSM internal project USM No.~11.15.41.

\end{document}